# Novel data and a new parameterization of the electric dipole strength in nuclei with 88 < A < 116.

**Eckart Grosse** [1]

*Institute for Radiation Physics,*
*Forschungszentrum Dresden-Rossendorf, 01314 Dresden, Germany, and*
*Institute for Nuclear and Particle Physics*
*Technische Universität Dresden, 01062 Dresden, Germany*
*E-mail:* `e.grosse@fzd.de`

**Arnd R. Junghans, Krasimir Kosev, Gencho Rusev,**
**Klaus D. Schilling, Ronald Schwengner and Andreas Wagner**

*Institute for Radiation Physics,*
*Forschungszentrum Dresden-Rossendorf, 01314 Dresden, Germany*
*E-mail:* `A.Junghans@fzd.de`, `k.kosev@fzd.de`, `rusev@tunl.duke.edu`,
`k.d.schilling@fzd.de`, `r.schwengner@fzd.de`, `a.wagner@fzd.de`

**Abstract**

A hitherto unexplored method for the experimental determination of the photon strength function up to the neutron separation energy was developed at the Radiation Source ELBE in Dresden. It was applied to various heavy nuclei, preferentially to nuclides with increasing distance to the N=50 neutron shell, and it covers the high level density excitation energy range above 4 MeV. The observed quasi-continuous spectra of scattered photons can be – after a proper correction for multi-step processes – directly combined to nuclear photo effect data from literature. A remarkably good match of the photon strengths as measured below and above the neutron emission threshold is observed. The wide energy coverage of the combined data forms an excellent basis to derive a parameterization for the dipole strength function fully covering the range across the nucleon separation energies. In addition to the parameters defining the deformation of the nuclear ground states only one additional constant is needed to describe the dipole strength in the nuclei with 88<A<116. The new parameterization differs significantly from the prescriptions generally used in network calculations, e.g. those of interest for the cosmic nucleosynthesis.

___________________

[1]  Speaker

## 1. Introduction

Electromagnetic radiation is a fundamental probe for the understanding of nuclei – similar to other systems of the micro-cosmos. Photo-nuclear processes were among the first nuclear reactions studied [1] and their appreciable strength has triggered the conclusion [2-4] that they are likely to play an important role for the cosmic nucleosynthesis: In the intense photon flux during high temperature cosmic scenarios particle emission thresholds are reached leading to the photo-disintegration of previously formed heavier nuclides. This so-called p-process is considered the main origin of more than 30 neutron-deficient nuclides [4-7] which cannot be produced in neutron capture reactions [2]. As shown previously [8], details of the dipole strength (here in n-rich nuclei) affect significantly the r-process path and even s-process branchings depend on nuclear excitations [6,9] induced by thermal photons. In contrast to the rather detailed knowledge [10] of the photon strength in the isovector giant dipole resonance (GDR) region well above the particle separation energies the corresponding information for the excitation region below is surprisingly scarce and by no means complete. Astrophysical network calculations have to account for the photon strength over the full range up to beyond particle thresholds as nucleosynthesis processes require high temperature and previously formed nuclei can be disintegrated in the resulting high photon flux.

The work described here is motivated by the observation [11, 12] that in many heavy nuclei the suggestive procedure of fitting a Lorentz-curve to the peak region of the GDR and extrapolating it to lower excitation energies fails to properly predict the strength close to the particle thresholds. Previous attempts to experimentally derive the strength down to energies far below the GDR have suffered from serious deficiencies: (a) in (n,γ) studies [12-14] only a narrow region is reached in the 1$^{st}$ step of the de-excitation cascade; (b) photon scattering experiments [15] neglecting the continuum part of the spectra can deliver correct strength information only up to limited excitation energy [16]; (c) experiments performed with He-ions [17] populate a diffuse distribution in spin. To properly assess observations [15, 18, 19] on additional 'pygmy' structures in photo-absorption data a proper knowledge of the underlying smooth strength function is urgently needed. Previous extrapolations to unstable nuclei as needed for nucleosynthesis calculations lacked sufficient experimental support and have largely been based on theoretical calculations [20, 21].

## 2. Photon scattering as a probe for the dipole strength

Experimental data for the complete photon strength covering a wide energy range were obtained by using photon scattering [22-24] in combination to published data on neutron photoproduction. As Mo isotopes have attracted much attention concerning their cosmic synthesis [7] a focus was put on the series of stable even Mo-isotopes [23]. Due to their $0^+$ ground states photon absorption experiments reach predominantly J=1-levels via their electromagnetic dipole strength. The analysis of the data is performed by carefully accounting for non-nuclear contributions [26] to the scattering and by extended investigations on the influence on inelastic scattering [22, 23]. As will be shown in the following, this hitherto unexplored method yields a strength function at clear variance to previous work [12, 13].

Experimentally, the photon strength function $f_\lambda(E)$ [18] in a region of excitation energy $E$ and spin $J$ is related to average ground state photon width $\overline{\Gamma_0}$ and average photo-absorption cross section $\overline{\sigma_\gamma}$:

$$f_\lambda(E,J) = \frac{\overline{\Gamma_0(E,\lambda)}}{E^{2\lambda+1} \cdot D_\lambda(E)} = \frac{2J_0+1}{(\pi\hbar c)^2 \cdot (2J+1)} \cdot \frac{\overline{\sigma_\gamma(E,\lambda)}}{E^{2\lambda-1}} \quad (1)$$

The average is formed over many levels, i.e. narrow resonances, as excited by photons with multipolarity $\lambda$ from the ground state spin $J_0$. Radiation with multipolarity $\lambda > 1$ contributes to the absorption only weakly such that $f(E) = f_1(E) + f_2(E) + ... \approx f_1(E)$. Close to the separation energies the average level distance $D_\lambda$ is small as compared to the experimental resolution resulting in a smooth energy dependence of the absorption [27-29]. As nuclear absorption is not easily distinguishable from the generally stronger non-nuclear processes, experiments set to directly determine it through extinction are extremely difficult [26].

The experiments described here determine the photo-absorption strength indirectly from elastic photon scattering observed using a bremsstrahlung continuum produced with the new superconducting electron linac ELBE [24] at Dresden-Rossendorf. The applied experimental method [22, 23] is innovative in its way of separating nuclear from non-nuclear processes and in its procedure of determining the ratio of elastic to inelastic scattering. Below the respective particle emission thresholds all nuclear levels with sufficient transition strength to the ground state are observed in elastic scattering, and the signal from an elastically scattered photon identifies their excitation energy. Inelastic scattering to levels other than the ground state can be corrected for in the data analysis – as will be discussed below. The use of large volume Ge-detectors in combination with escape-suppression shields is very important as this combination leads to a nearly diagonal detector response-matrix. The well shielded set-up and the high energy resolution (5-8 keV FWHM for 4-9 MeV) of the scattered photons allowed the clear identification and suppression of ambient background. For the even Mo-isotopes several grams of isotopically enriched material were irradiated under very similar experimental conditions at an electron beam energy of 13.2 MeV. By observing the scattered photons at 127° with respect to the collimated photon beam [22, 23] angular distribution corrections for dipole transitions are small and quadrupole transitions are suppressed.

As can be seen in Fig. 1, the intensity rapidly fades away with increasing energy after crossing the neutron emission threshold since then photon absorption nearly completely results in nucleon emission. In the case of $^{92}$Mo the photon scattering above 9 MeV is strongly influenced by the opening of the ($\gamma$, p) channel at 7.5 MeV; in the experiments at ELBE this channel was adressed [29] in an activation study, partly by using a fast pneumatic transport system. In all investigated nuclei an important contribution to the spectra stems from non-nuclear scattering. Non-nuclear processes can be simulated [22, 23, 26] to high accuracy and to yield the "true" intensity $I_\gamma$ (originating from nuclear scattering) to be used in eq. 2 their contribution has to be subtracted from the spectra. A test of the accuracy of the subtraction procedure is obtained from the experimental data directly [22, 23]: Due to the difference of the neutron binding energies of $^{98}$Mo and $^{100}$Mo a comparison of the two spectra between the neutron thresholds $S_n(^{100}$Mo$)$ and $S_n(^{98}$Mo$)$ yields information about the non-nuclear background - under the suggestive assumption that atomic cross sections are identical for both isotopes.

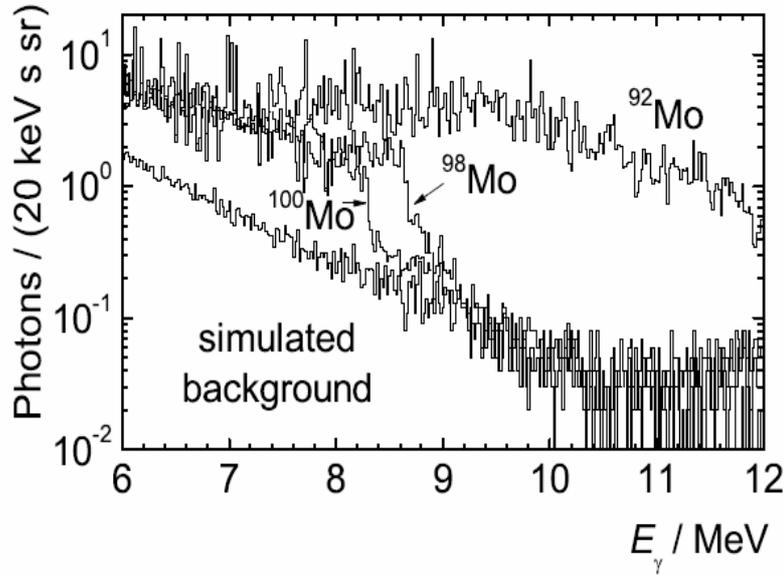

Fig. 1: Photon scattering rate as observed in the irradiation of $^{92}$Mo, compared to a Monte-Carlo simulation of the non-nuclear background. At the neutron thresholds for $^{98}$Mo and $^{100}$Mo the photon scattering yield decreases suddenly. For $^{92}$Mo this threshold is much higher at 12.7 MeV.

The high level density in combination with Porter-Thomas fluctuations [25] causes a large portion of the strength to appear in a quasi-continuum of not resolved peaks. Therefore, the photon strength in a given energy interval has to be obtained by integrating the complete spectra after subtraction of the non-nuclear background. As is seen in Fig. 1, the data in the last MeV below $S_n$ are dominated by the continuum to at least 70 %. An analysis of resolved peaks alone [15] only delivers information on the excitation energies of the configurations with localized strength and leads to a systematic underestimate of the total strength.

A further correction to be applied to the data accounts for inelastic scattering, i.e. transitions branching to lower lying excited states and their feeding. A Monte-Carlo-based method was derived to obtain an accurate correction for this from the data directly. The simulations follow a procedure used in a respective analysis [13, 14] of (n,γ)-data. They are made to obtain approximations for the feeding contribution $I_{feed}$ to the intensity $I_\gamma$ and for $\sum \Gamma_i / \Gamma_0$ (as describing the branching) and are performed on the basis of a random matrix ansatz: The nearest neighbour distance distributions of the levels are assumed to be of Wigner type and their transition widths should follow Porter-Thomas [25] distributions. On the basis of these assumptions, the Axel-Brink hypothesis [28, 29] and a level density systematics [27] random level and decay schemes were created. Starting from an arbitrary first input parameterization for $f_1$, a correction for feeding and branching was attained by calculating and inserting $I_{feed}$ and $\sum \Gamma_i / \Gamma_0$ into:

$$f_1(E) = \frac{\overline{I_\gamma} - \sum_{f.t.} \overline{I_{feed}}}{3(\pi \hbar c)^2 \cdot E \cdot \Phi(E) \cdot n_t} \cdot \frac{\sum \Gamma_i}{\Gamma_0}(E) \qquad (2)$$

The rate $\bar{I}_\gamma$ of photons scattered from the $n_t$ target nuclei was averaged in a bin around the energy $E$ and the sum in the 2$^{nd}$ term has to be taken over all transitions (*f.t.*) feeding this interval; it is also used for averaging the width ratio of all branching transitions $\Gamma_i$ vs. the decay width $\Gamma_0$ to the ground state. A self-consistent solution for $f_1$ was found iteratively by comparing the input $f_1$ to the result from (2) and minimizing $\chi^2$. The photon flux $\Phi$ was determined [22, 23] by simultaneously measuring elastic scattering intensities originating from $^{11}$B added to the target material.

### 3.   Combined analysis of photon scattering and photo-neutron data

The obtained dipole strength functions $f_1$ show a rather smooth dependence on energy up to $S_n$ and match very well the ($\gamma$,xn)-data [30, 31] directly above $S_n$; see Fig. 2. Thus it is justified to combine the two data sets to extract a continuous dipole strength function. We attempt to parameterize it by a Lorentzian tailing down from the GDR. As was shown theoretically [33] a description of the E1 photon absorption cross section by a Lorentzian is appropriate albeit the total width $\Gamma$ of a GDR in a heavy nucleus is dominated by spreading [30, 32] and not by escape, i.e. direct decay. This width describes the spreading due to the coupling of the GDR to the underlying continuum of p-h and more complicated configurations [32]. The wide range of excitation energy spanned by the combined data leads to a high sensitivity for a possible energy dependence of the total width $\Gamma(E)$ of the Lorentzian to be compared to:

$$f_1(E) = \frac{2\alpha \cdot S_{TRK}}{9\pi \cdot (\pi\hbar c)^2} \cdot \sum_{k=1,2,3} \frac{E \cdot \Gamma(E)}{(E_k^2 - E^2)^2 + E^2 \cdot \Gamma(E)^2} \qquad (3)$$

To treat the influence of the nuclear shape separately from the spreading we explicitly account for it by summing the GDR components $E_1$, $E_2$ and $E_3$ resulting from the nuclear deformation [34, 35] specified by the parameters $\beta$ and $\gamma$. Using the eccentricity $\varepsilon \approx 0.95\,\beta$ the resonance energies for the three components k = 1,2,3 are given by [35]

$$E_k = E_o \cdot \left[1 - \frac{2}{3} \cdot \varepsilon \cdot \cos(\gamma + \frac{2k}{3} \cdot \pi)\right] \qquad (4)$$

The parameter $\alpha$ measures the conformance of the integrated E1 strength to the model independent Thomas-Reiche-Kuhn (TRK) sum-rule [10], requiring

$$S_{TRK} = \int_0^\infty \sigma_\gamma(E)\,dE = 60\,NZ/A \text{ MeV mb} \qquad (5)$$

It was observed [30, 34, 36] that a 'standard' of $\Gamma(E_{GDR}) = \Gamma_S = 4$ MeV describes the GDR shape in all non-deformed nuclei in the mass range 80 < A < 120 reasonably well. Although in many heavy nuclei, especially in $^{100}$Mo, the apparent width of the GDR is exceeding 4 MeV considerably, the use of $\Gamma = 4$ MeV in three overlapping Lorentzians shifted by the amount as given by (4) leads to a perfect fit of the GDR region (cf. Fig. 2). In the case of Mo the deformation parameters were experimentally determined by Coulomb excitation [37] in agreement to theoretical results [38]. Obviously the large apparent width is mainly due to the split of the GDR as caused by the axial deformation $\beta$, with the non-axiality $\gamma$ inducing an additional change of the shape near the centre of the GDR only (cf. Fig. 4).

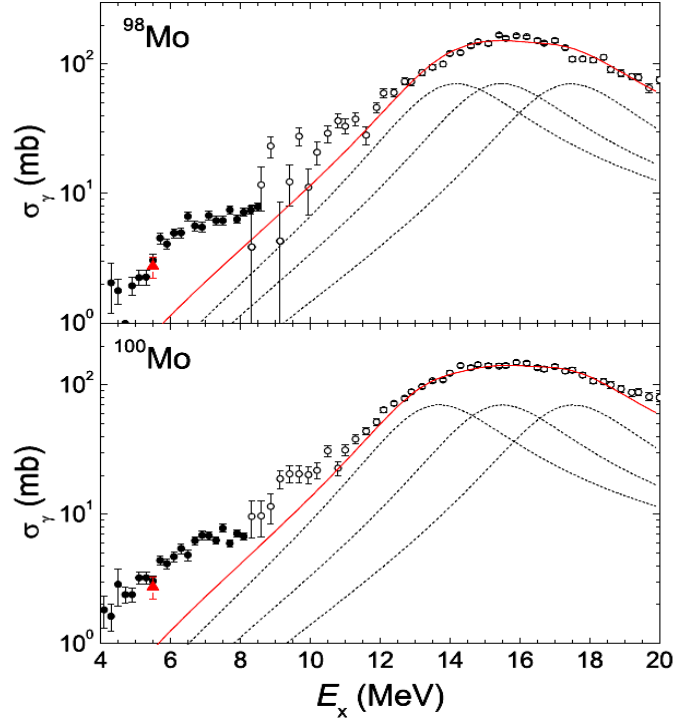

Fig. 2: Experimental dipole absorption cross section for $^{98,100}$Mo. The data from the photon scattering (●) are combined to (γ,n)-data (o) [31]. A $^{98}$Mo(n,γ)-data point (red ▲) [12] as calculated from the strength function at the respective energy is inserted. The lines depict the parameterization corresponding to δ=0, α = 1, β= 0.19 and its decomposition into the components belonging to the three principal axes for γ =32°.

Far off resonance the dipole strength (3) is proportional to Γ. We parameterize the energy dependence of Γ by

$$\Gamma(E) = \Gamma_S \cdot (E/E_i)^\delta$$

with δ as parameter to be defined by a fit to the combined data. Especially from the slopes far below the resonance energy – as obtained from the scattering data and confirmed by a $^{98}$Mo(n,γ)-data point [12] as compared to both of the Mo isotopes – we obtain δ = 0.0 (0.1) and thus <u>no</u> gamma-energy dependence of the spreading width Γ. This is clearly at variance to previous investigations for nuclei in this mass range [12] and a more detailed discussion of this point will be given in the next paragraph. A good fit to the data is obtained with α = 1 (with an accuracy of 0.13) for $^{88}$Sr [22] and the Mo isotopes investigated [23]. In the calculations a weak M1 contribution was included as derived from previous work [38].

A comparison of the combined data for the three Mo isotopes with the smooth Lorentzian curve from (3) as plotted in Fig. 2 shows a slight enhancement of the strength in a flat maximum at ~ 9 MeV. In $^{100}$Mo this structure appears above $S_n$ and in $^{98}$Mo the region directly above $S_n$ = 8.6 MeV irregularities are indicated despite large statistical uncertainties. Earlier, a cross section enhancement at 9.1 MeV was found [39] in $^{90}$Zr and most of that strength was shown not to be M1 [40]. This result as well as the current Mo data may be related to broad resonance-like structures recently observed in $^{130}$Sn and $^{132}$Sn [19] at an energy just above 9 MeV. Ongoing shell model calculations [41] combined with collective excitations and a Hartree-Fock or Random Phase Approximation should clarify how important multi-particle-hole configurations with $J^\pi = 1^-$ are for extra dipole strength outside of the GDR.

## 4.    Comparison to previous strength function studies

The parameterization derived here is clearly at variance to an earlier method [12] widely used to describe data below the particle emission thresholds. The main difference is that for not too strongly deformed nuclei the previous approach described spreading and deformation induced widening of the GDR by a common width. The resulting extra strength at low energies was compensated by an energy dependence parameter $\delta = 2$ inserted into $\Gamma(E) = \Gamma_0 \cdot (E/E_0)^\delta$. In contrast we calculate a dipole strength distribution with $\delta = 0$ inserted into (3) to (6) by strictly observing a clear-cut separation of the two causes for the widening of the GDR and by using standard deformation parameters [42, 43]. The absorption cross section in the GDR as well as below the neutron threshold is then given by (1), in agreement to the data for several Mo isotopes, as described above. To show the influence of our choice to other nuclei in the A = 100 range we present in the following results for various nuclei which have previously been discussed using the differing ansatz with $\delta = 2$ for the total width of the GDR [12].

The GDR in $^{90}$Zr was studied not only by $(\gamma,xn)$ [36, 44] but also through $^{89}$Y$(p,\gamma)$ [45], populating not only the ground state of $^{90}$Zr. In Fig. 3 we show from that experiment the strength function to the first excited state – shifted by its excitation energy. Within the GDR range it agrees quantitatively to the $(\gamma,n)$-data, which – of course – are connected to the ground state. As stated [45], this fact constitutes an impressive test of the Axel-Brink hypothesis [28, 46]. Additionally, these data show a perfect agreement to a single Lorentzian with $\Gamma = 4$ MeV, $\alpha = 1$ and $\delta = 0$. This is in full accordance to our findings [23] in the neighbouring Mo-isotopes, as described above. Presently we are investigating $^{90}$Zr at ELBE with the methods developed in the study of the Mo-isotopes and first results [22] agree to what is shown in Fig 3.

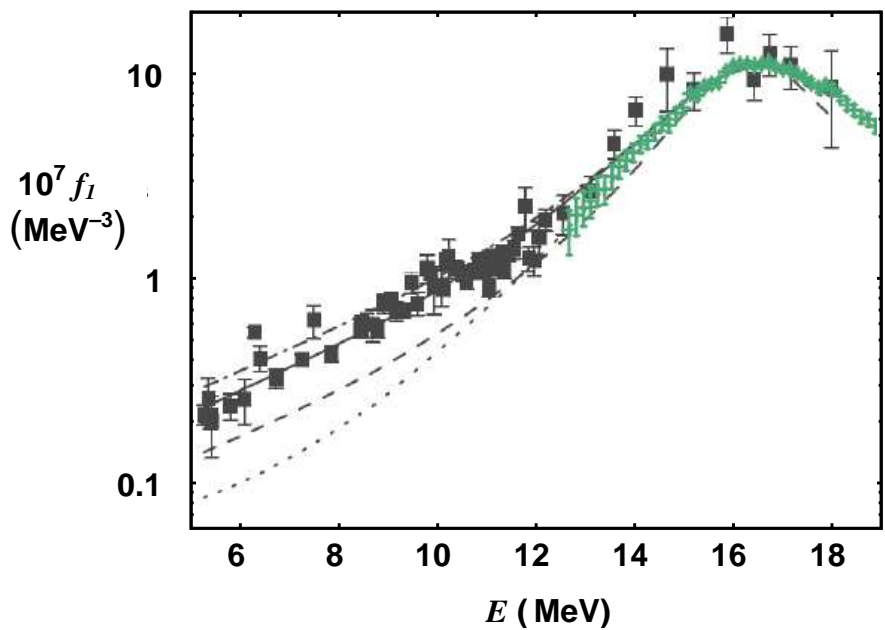

Fig. 3: Experimental dipole strength functions for $^{90}$Zr in comparison to various parameterizations. The full line corresponds to $\Gamma=4$ MeV and $\delta=0$, $\alpha = 1$, $\beta = 0$. The green crosses stem from the $(\gamma,n)$-experiment and the black data points are data obtained by $^{89}$Y$(p,\gamma)$ to the 1$^{st}$ excited state in $^{90}$Zr [45, quoted in 11]. As seen from the broken lines in the Figure, there have been less successful attempts to describe the data by other parameterizations [11, details can be found there].

A nucleus extensively studied in (n,γ) with respect to its dipole strength function is $^{106}$Pd [12]. In Fig. 4 we compare these data to the results from $^{nat}$Pd(γ,n) and to results of our parameterization, based on β= 0.19, Γ = 4 MeV, α = 1 and δ = 0, 1 and 2. Apparently the (n,γ)-data favour 0<δ<1. The deformation influence of the triaxiality is only seen near the top of the GDR and not in the low energy tail. According to (3) this holds for triaxially deformed nuclei in general.

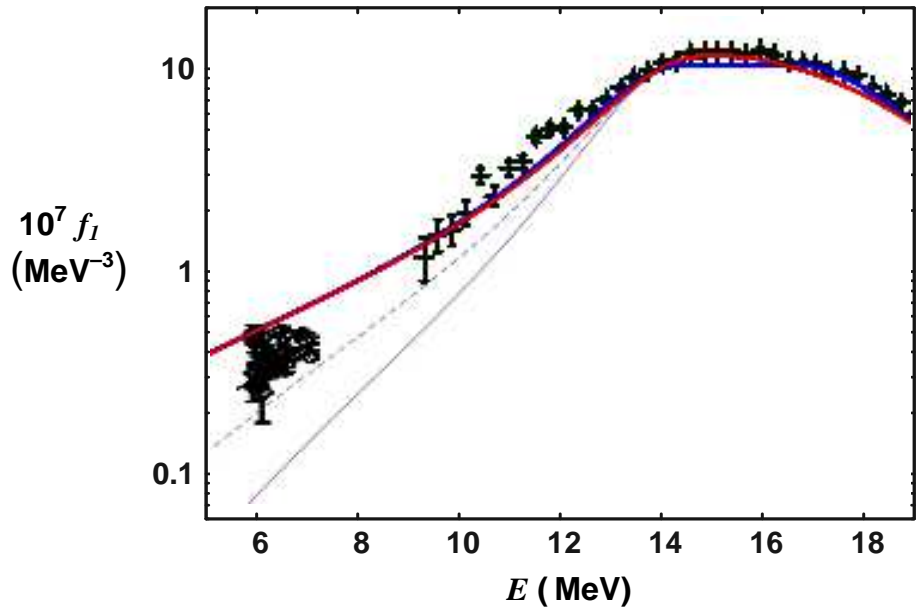

Fig. 4: Experimental dipole strength functions for $^{106}$Pd in comparison to various parameterizations: The full lines correspond to δ=0, α = 1, β= 0.19; in red with a triaxiality γ = 30° and in blue for γ = 0. The broken line represents δ = 1 and the lower thin line depicts δ = 2.
The green crosses stem from $^{nat}$Pd(γ,n) and the black data points (♦) are data obtained by $^{105}$Pd(n,γ).

It is interesting to apply the outlined approach to odd nuclei as well. A detailed study of two-step gamma-ray cascades was performed [47] following the neutron capture in $^{107}$Ag in several s-wave resonances. To yield strength function on an absolute scale these data have to be normalized to thermal capture intensities, average photon widths and level densities, in this case for $J^π = 1^-$. The respective procedure used in ref. [47] was repeated on the basis of the most recent Nuclear Data Sheets for A = 108 [48] and previous very detailed gamma decay data [49]. From this information, apparently not used by [47], it appears that the normalization factor 91/137 = 0.66 used in [47] is to be replaced by 1. As in [49] all transitions following the n-capture are properly normalized to the final nucleus' radioactivity the additional normalization is to be omitted. After this modification a combination of the (n,γ)-results to the (γ,n)-excitation function [36, 44] yields a smooth energy dependence similar to the slope indicated in the capture data. An additional increase (by a factor of 1.7) in the strength function from [47] would result, if the s-wave average level spacing $D_0$ used there in eq. (2) would be replaced by the value from a recent compilation [50]. With or without such an additional increase the combined data are reasonably well described by the parameterization without an energy dependent width – as can be seen in Fig. 3. The moderate deformation β = 0.15 is in agreement to spectroscopic information [42, 43] and the accordance to the TRK sum is demonstrated by α = 1.

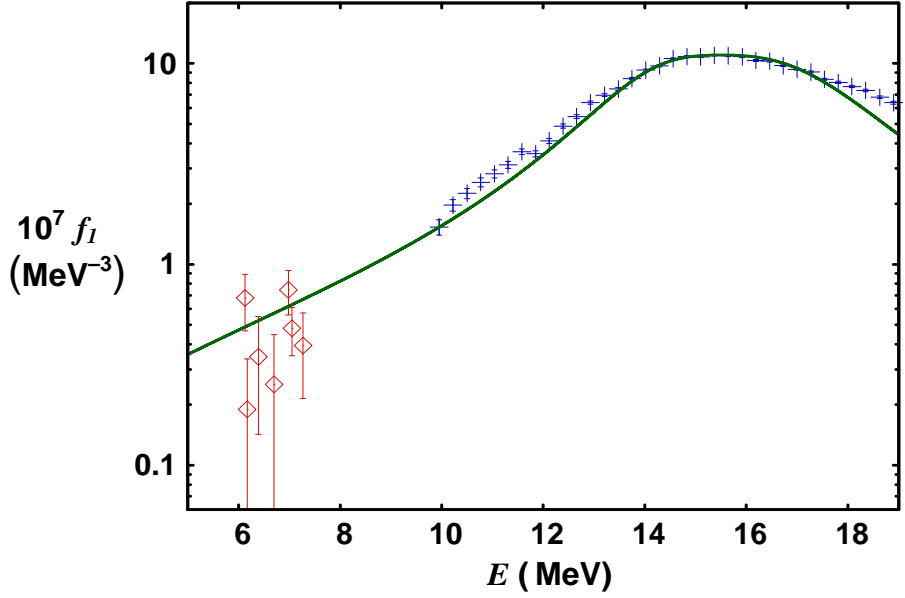

Fig. 5: Experimental photon strength function for Ag. Data from (n,γ) [44] are combined to (γ,n)-data [47, divided by 0.66, see text]. The line depicts the parameterization with δ=0, α = 1 and β= 0.15, γ =17°.

In $^{115}$In the dipole strength in the GDR [36, 44] and that from (n,γ) [12] are well parameterized by δ=0, α=1 and a combination of Lorentzians of width Γ = 4 MeV, when a ground state deformation β = 0.13 from spectroscopic data [42, 43] and γ = 17° are introduced into (3) and (4).

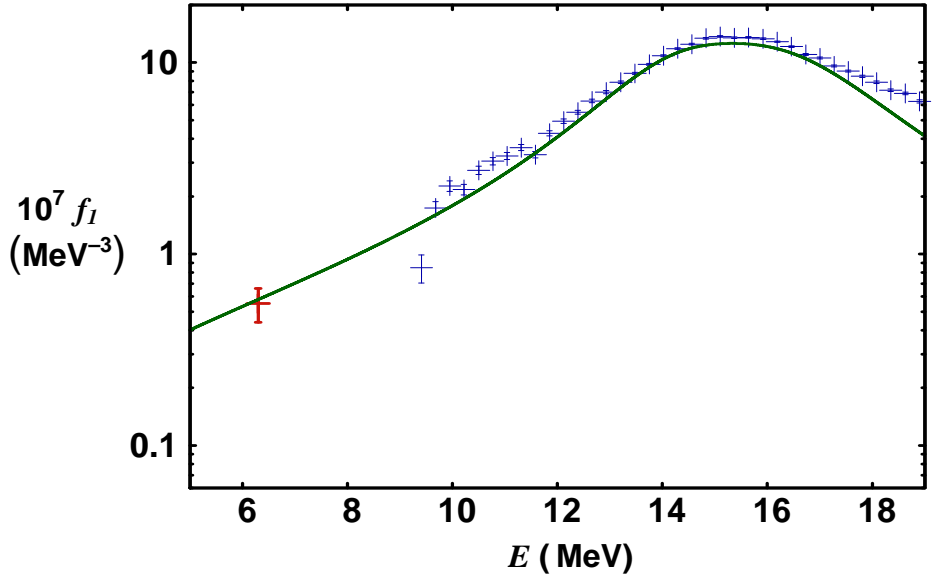

Fig. 6: Experimental photon strength function for In. Data from (n,γ) [44] are combined to (γ,n)-data [12]. The line depicts the parameterization corresponding to δ=0, α = 1 and the deformations β= 0.13, γ =17°.

Additional data [51] from photon scattering experiments confirm the results presented here and further analysis is under way to extend the present study to heavier nuclei. Eventually beams of monochromatic photons will allow more detailed investigations in the strength function field and the new intense sources of fast neutrons promise interesting complementary data.

## 5. Conclusion

The response of nuclei to electromagnetic radiation can well be studied by photon scattering using bremsstrahlung. For the data presented in this work a sophisticated analysis of the quasi-continuum part of the cross section was performed after a detailed determination of non-nuclear contributions. Contributions due to Rayleigh and Delbrück scattering are negligible at backward angles for $E_\gamma$ > 4 MeV and our investigations are limited to that range. Thus they do not allow to make statements on the dipole strength below 4 MeV although distinct theoretical predictions [52] were made there and they are not giving information on the enhanced M1 strength caused by the scissors mode. Using a sufficiently high endpoint energy and correcting the data for inelastic processes allows to directly combine the dipole strength functions obtained from (γ,γ)-data to those resulting from photonuclear experiments like (γ,n). In the cases studied these two data sets match very well and the resulting combined dipole strength function can be considered appropriate for the full range from low energy up to the GDR.

A line shape based on a Lorentzian for each axis corresponding to the ground state deformation parameters fits these data very well. As it turned out it is mainly β which is of importance and the respective information is easily available from nuclear quadrupole moments [42] and B(E2)-values [43]. An important component of the parameterization as proposed here is a strict separation of the GDR 'breadth' into a part resulting from the deformation and one caused by the spreading into the multi-particle multi-hole configurations underlying the giant resonance. This spreading is well accounted for by using Γ = 4 MeV for nuclei with A ~ 100 [36] and with this choice no strong deviation from the TRK sum-rule is observed. It should be noted here that we have made no attempt to include vibrational excitations of E2-type in our analysis. Their much smaller oscillation frequencies in comparison to those of the GDR suggest an adiabatic situation with possible E1×E2-strength appearing at energies above the GDR. Such an up-shift is in accordance to the Axel-Brink hypothesis and an inspection of the experimental data shows that there is room above the GDR for some increase of the strength function prediction.

An energy <u>in</u>dependence of the spreading width is indicated by the photon energy dependence of the scattering data. As shown here they are in agreement to data from neutron capture by e.g. $^{98}$Mo, $^{107}$Ag and $^{115}$In (with the case of $^{105}$Pd(n,γ) being less clear, cf. Fig. 4). Strong additional support to our parameterization – which is at variance to prescriptions widely used previously [11,12] for the interpretation of photon accompanied processes – comes from the (γ,p)-data in $^{90}$Zr as presented in Fig. 3. Together with a prediction for the GDR energy and for the ground state deformation our ansatz delivers an expression of possible use when no GDR-data are available, which may be of importance for calculations including nuclei far off stability participating in nucleosynthesis processes.

**Acknowledgements**

P. Michel and the ELBE-Crew made the investigations possible by delivering optimum beams from the Radiation Source ELBE. R. Beyer, M. Erhard, A. Hartmann, C. Nair, N. Nankov and W. Schulze provided very valuable support during the difficult experiments. Of major importance for this research was the availability of the EXFOR database [36] and thanks are due to V. Varlamov for useful communications. Intensive discussions with H. W. Barz, F. Becvar, F. Dönau, S. Frauendorf, M. Krticka, H. Lenske, N. Tsoneva and R. Wünsch are gratefully acknowledged. The DFG has funded parts of this work under the contracts Gr1674/1-2 and Do466/1-2.